%Paper: hep-ph/9307373
%From: orloff@surya11.cern.ch (Jean Orloff)
%Date: Fri, 30 Jul 93 20:29:09 +0200

\input harvmac.tex
\input psbox.tex
%2345678 1 2345678 2 2345678 3 2345678 4 2345678 5 2345678 6 2345678 7 2345678
% -----------------------------------RefERENCES
\def\PRL{ {\sl Phys. Rev. Lett.}   }

\def\NP { {\sl Nucl. Phys.}        }

\def\PL { {\sl Phys. Lett.}        }
\def\PH { {\sl Physica}		}
\def\ie{{\it i.e.}}

\Title{\vbox{\hbox{CERN-TH 6925}\hbox{IEM-FT-74/93}}}
{Effective Operators and Extended Symmetry}

\centerline{J.-M. Fr\`ere\footnote{$^\dagger$}{Ma\^{\i}tre
de Recherche FNRS, Corresponding Fellow, CERN.},
M. Tytgat\footnote{$^\flat$}{Aspirant FNRS.}}
\centerline{Service de Physique Th\'eorique}
\centerline{Universit\'e Libre de Bruxelles, CP 225}
\centerline{Boulevard du Triomphe, B-1050 Bruxelles, Belgium}
\bigskip
\centerline{J.M. Moreno}
\centerline{Inst. de Estructura de la Materia}
\centerline{Serrano 123,  E-28006 Madrid, Spain}
\bigskip
\centerline{J. Orloff\footnote{$^\sharp$}{e-mail: orloff@dxcern.cern.ch}}
\centerline{Theory Division, CERN}
\centerline{CH--1211 Geneva 23, Switzerland}
\vskip .3in

In this note we expand on our previous study of the implications of LEP1
results for future colliders. We extend the effective operator-based
analysis of De R\'ujula et al. to a larger symmetry group, and show
at which cost their expectations can be relaxed. Of particular interest to
experiment is  a rephrasing of our previous results in terms of the Renard et
al.  parametrization for the gauge boson self-couplings (slightly extended to
include $\delta g_{\gamma}$). We suggest the use of a ($\delta g_{\gamma}$,
$\delta g_{Z}$) plot to confront the expectations of various models.

\vfil
\noindent CERN-TH.6925/93
\Date{07/93}

\newsec{Outline}

The purpose of the present paper is to investigate to which extent the extra
degrees of freedom ---possibly associated with light particles--- resulting
from  an extended symmetry may alleviate the constraints on future high energy
experiments resulting from the high precision measurements  of LEP1 at the $Z$
peak.

In a previous note \ref \us{J-M. Fr\`ere, J.M. Moreno, J. Orloff and M.
Tytgat,  \PL {\bf B292}  (1992) 348.}, we outlined the main characteristics of
a  minimal extension of the type $SU(2) \times U(1) \times U(1)' $, and
reached  the tentative conclusion that the presence of a {\bf light} $Z'$
boson, with  anomalous, gauge-invariant, dimension-6 couplings could remain
undetected at  LEP1 while providing,  at least in terms of statistics,
significant  contributions already at the moderately higher energies to be
reached by LEP2.   In a way, this constitutes a loophole in the argument of
\ref\deRuja{A. De  R\'ujula, B. Gavela, P. Hern\'andez and E.~Mass\'o,
Nucl.Phys. {\bf B384}  (1992) 3.}, to the extent that some of the implicit
conditions of \deRuja\ are  not met; we acknowledge however quite willingly
that the construction of this  counterexample is involved enough to underline
the intrinsic power of the  general arguments\foot{at least those based on
tree level considerations}  developed in \deRuja.

We return to our previous analysis, with the purpose of expliciting somewhat
the construction and providing our experimentalist colleagues with the basic
tools to confront our approach with their Monte-Carlo simulations. In
particular, we study the impact of our model on $M_W$ and LEP1 data like
$(e_A,e_V)$, which determines the extreme values allowed for our parameters
$\lambda$, the ratio of the $U(1)'$ to the standard $U(1)$ couplings, and
$\varepsilon$, the strength of the  hypothetical anomalous coupling of the
$Z'$.  In addition to the angular  distribution of the produced W pairs, which
is not directly accessible to experiment, we establish the relation of the
extended model to the now usual parametrization of the amplitudes in terms of
the variables ($\delta g_Z$, $\delta \kappa_Z$, $\delta  \kappa_{\gamma}$)
introduced by  Renard et al.  (see for example \ref\renar{M. Bilenky, J.-L.
Kneur, F.M. Renard, D. Schildknecht, Bielefeld preprint, BI-TP-92-25 (1992)}),
which we need, however, to extend slightly by allowing for $\delta
g_{\gamma}$.  This reparametrization is not entirely  trivial, as the approach
of Renard et al. assumes that the $Z$ and $\gamma$ are the only neutral vector
bosons. The $Z'$ is thus simulated at the cost of introducing
momentum-dependent form factors, both for the $Z$ and the $\gamma$ anomalous
couplings.  Fortunately, at tree level (and we limit the study of the
extensions to their  tree-level effects) these form factors only depend upon
$\sqrt s$. In addition  to the formal expression of the Renard parameters in
terms of our fundamental  Lagrangian variables, we work out typical values, as
allowed by the latest LEP1  results \ref\cathy{C. De Clercq, ULB preprint
IISN0379-301X (1993),  to appear in {\it``Rencontres de Moriond"} proceedings
(Electroweak session, march 1993)}.

\newsec{General Considerations}

Effective Lagrangians provide a systematic way of parametrizing new physics
beyond the Standard Model (SM) \ref\weinberg{S. Weinberg, \PH {\bf A96} (1979)
327; H. Georgi, {\it Weak Interactions and Modern Particle Theory}
Benjamin-Cummings, Menlo Park, CA (1984).}. In such an approach, any new
physical input merely boils down to a set of numerical coefficients weighting
higher-dimensional (and hence non-renormalizable) operators, constructed in
terms of the low-energy fields alone. If the approach is to be of any use,
ways must  be found to truncate the infinite list of such operators down to a
small  subset.  An obvious guide is the dimensionality of the operators. If
indeed the  new physics is supposed to originate at some large scale
$\Lambda$, this scale  is expected to appear in the denominator of the
coefficient of each operator.

In gauge theories, the notion of gauge invariance of the operator basis has
been abundantly discussed in the literature. This requirement has important
consequences on the classification of operators according to dimension.

Let us consider indeed a typical example :
\eqn\eIi{O_{WB}^{naive}=W_{\mu\nu}^3 B^{\mu \nu}.}
This operator is apparently of dimension 4, although non-renormalizable, since
it corresponds to a non-minimal coupling of the $W$.  Quite obviously, this
operator also violates gauge invariance, and even the global $SU(2)$
invariance  of the effective Lagrangian.  It has been argued that the notion
of gauge  invariance is of little relevance, since any non-invariant operator
can be  written in the form of a gauge-invariant term, evaluated in some
unitary gauge  after symmetry breaking.  An obvious example to realize this
consists in  rewriting  $O_{WB}^{naive}$ as
\eqn\eIii{ O_{WB} = \phi^+ \sigma^i \phi W_{\mu \nu}^i B^{\mu \nu}.
}
This term is now obviously gauge-invariant, and coincides (up to a dimensional
coefficient) with the previous expression after symmetry breaking. This step
however involves a change in the dimension of the operator considered, and
stresses that it should, in a fully renormalizable framework, be penalized by
two powers of the high scale. While the requirement of gauge invariance of the
effective Lagrangian (before symmetry breaking) is not a mathematical
necessity, we thus observe that, on physical grounds, it is a useful guide in
tracking down the severity of the actual dependence upon the high scale.

The enforcement of gauge invariance of the operator set, and the restriction
to  operators of dimension $\leq 6$ , has allowed De R\' ujula et al.
\deRuja\ to establish powerful constraints on the possible observations at
future colliders. While we will return to these assumptions, we briefly
outline  our current understanding of the situation. The reason for these
constraints is  simple. Gauge invariance typically relates Feynman vertices
with different  numbers of gauge fields; for instance in the case of $O_{WB}$
above, terms with  external legs $BWW$ and $BW$ are automatically related.
Owing to its high  luminosity at the $Z$-peak LEP1 has brought strong
constraints on the masses and  couplings of the gauge bosons ---thus
constraining severely the 2-point  functions. As a result, operators leading
to both 2-gauge and 3-gauge boson  terms are severely constrained already at
the tree level. However, ``blind  directions'' also exist at tree level. In
particular, such is trivially the  case of operators which do not contain any
$2-W$ contributions, and the  archetypal example is:
\eqn\eIiii{O_W ={1\over 3!}\varepsilon_{ijk} W_\mu^{\nu j} W_\nu^{\lambda k}
  W_\lambda^{\mu i} .
}
In addition to the undisputable successes of the tree-level considerations,
several groups have studied further constraints arising from the inclusion of
the new operators in 1-loop diagrams\ref\loops{ P. Hernandez and F.J. Vegas,
\PL{} {\bf B307} (1993) 116-127;\hfil\break R. Escribano and E. Mass\'o, \PL{}
{\bf B301} (1993) 419.}, typically generating contributions to  low-energy
parameters measured to a high accuracy (for instance $(g-2)$, or  the electric
dipole moment of various leptons and quarks).  These studies show  that, in a
number of cases, tree-level blind directions can bring important  1-loop
contributions, and can therefore be partly excluded. While these
considerations are highly plausible, it should of course be kept in mind that
the same high-energy contributions, which are responsible for anomalous W
operators, can also -and in general do- generate contact terms, providing
direct contributions to these parameters. Such terms are furthermore often
required to regulate the divergences due to the inclusion of effective
higher-dimension operators in loops. The presence of such uncontrolled
counterterms thus precludes in general firm conclusions based on loop
contributions alone.

The situation may appear gloomy for some of the often advertised purposes of
future accelerators. We want however to emphasize that any effective
Lagrangian approach assumes in some way the existence of a ``desert'' between
the exactly treated ``known'' low-energy scale, and the high-energy scale,
described only by effective operators. That the harvest of new interactions in
such a desert be meagre, should then not be too surprising.

Reducing the scale at which new physics sets in can be treated in two ways.
The  most obvious one is to to include higher-dimensional operators in the
analysis; this however quickly makes the model completely unpredictable, since
there is a large number of operators beyond dimension 6. The other way to see
things is to insist right away on the presence of further particles in the
basic spectrum (think for instance of a possible $Z'$). Here also, the
previous approach has to be reviewed. Even if only dimension-6 operators are
constructed, these must involve new fields and, correspondingly, more degrees
of freedom are injected.  Actually, this may be seen as a curious twist on one
of the basic assumptions usually made. It is customary indeed to assume {\bf
only} the minimal $SU(2)\times U(1)$ symmetry to classify the additional
effective operators. At first sight, this seems to be a liberal assumption,
since any {\bf larger} symmetry is expected to introduce {\bf more}
constraints on the form of the Lagrangian. The argument fails, however, if the
extra symmetry considered, as is often the case, imposes the presence of an
enriched intermediate-energy  particle spectrum, allowing us to write down a
much  larger set of operators.

We are interested in building such an example, just to check how much our
expectations for physics at new accelerators would be affected. We present
here in some detail the simplest case we could think of, namely the extension
of the original symmetry group to $SU(2) \times U(1) \times U(1)'$, with a
relatively light $Z'$. We note in passing that the first $U(1)$ being
unchanged, the gauge boson associated with $U(1)'$ is orthogonal to the
photon,  which reduces from the onset any possible consequences on low energy
anomalous  magnetic or eletric dipole moments.

Our purpose in this note is to go back to the fundamental assumptions, and to
examine from the start whether any escape from the above-sketched analysis is
indeed possible.

\newsec{The Model}

We thus proceed in two steps, as in \us. First, we add the extra
Abelian gauge boson $B'$, together with a scalar $\chi$ providing for its
mass. Right-handed neutrinos ($\nu_R$) also need to be introduced,  which,
except for the cancellation of anomalies, play no role in our considerations.
The matter content and the additional hypercharges are summarized in Table 1.
Two more parameters are needed to determine the modified neutral current
Lagrangian: the strength of the new interaction ($g_1'$) and the mass of the
new gauge boson ($M_{Z'}$). In this minimal model, we have fixed the scalar
content and as a result the extra mixing angle ($\theta_3$) between $Z$ and
$Z'$ is  not an independent parameter. For the couplings, it is convenient to
turn to the variables $e$, $\theta_W$ and $\lambda$   ($e = g_2 s_W = g_1
c_W$, $\lambda = s_W g_1'/g_1$).

But this extra symmetry might not be the end of physics, and we next
parametrize our further ignorance by an effective operator approach. A typical
new dimension-6 operator is obtained by replacing the $B$ of $O_{WB}$ by a
$B'$ field, yielding
\eqn\eIiv{{\cal L}_{W B'}\doteq {\varepsilon \over v^2} O_{W B^\prime}
  = {\varepsilon \over v^2} \phi^+ \sigma^i \phi W_{\mu \nu}^i
     B^{\prime \mu \nu}.
}
With the introduction of the new operator $O_{WB'}$ and after S.B., the
bilinear part of the Lagrangian for the neutral gauge bosons becomes:
\eqn\eIv{
{\cal L}_{NGB} = -{1\over4} W_{\mu \nu} {\cal K} W^{\mu \nu}
  + {1 \over 2} W_{\mu} {\cal M}^2 W^{\mu },
}
where
\eqn\eIvi{\eqalign{
W_{\mu \nu}&= (\hat W_{\mu \nu}^3 \ \ B_{\mu \nu} \ \ B'_{\mu \nu} ),\cr
\hat W_{\mu \nu}^3 &=\partial_{\mu}W^3_{\nu} -
  \partial_{\nu}W^3_{\mu}\cr}
}
with\foot{Notice that the use of $v^2$ (instead of a higher energy scale
$\Lambda^2$) as a dimensionalizer in \eIiv{} reflects our phenomenological,
low-energy approach. With this choice, an $|\varepsilon|\ll \varepsilon_{crit}
=1$ has small effects, $\varepsilon_{crit}$ being the extreme value that makes
${\cal K}$ degenerate. We do not wish to address the question of the origin of
this $\varepsilon$ here.}
\eqn\new{{\cal K} =\pmatrix{1 & 0 & \varepsilon \cr
                                    0 & 1 &   0 \cr
                    \varepsilon & 0 & 1\cr}
}
and
\eqn\newi{{\cal M}^2 = \pmatrix{{1 \over 4} g_2^2 v^2 &- {1 \over 4}
 g_2 g_1 v^2 & {1 \over 5} g_2 g_1' v^2 \cr
                                    -{1 \over 4} g_2 g_1 v^2 & {1 \over 4}
  g_1^2 v^2 &   -{1 \over 5} g_1 g_1' v^2\cr
   {1 \over 5} g_2 g_1' v^2 & -{1 \over 5} g_1 g_1' v^2 & {4
\over 25} g_1'^2 v^2 +  g_1'^2 V^2\cr}
}

We need to  diagonalize simultaneously the mass and the kinetic matrices for
neutral fields, ending up with the canonical form for the latter. In addition
to rotations in $W_{\mu}$--space, this involves a rescaling of the fields. We
thus define $\cal S$ relating the physical neutral fields to the original
gauge fields\foot{This redefinition matrix ${\cal S}$ is  {\it not} unitary.}
by:
\eqn\eIvii{\eqalign{
{\cal S}^t {\cal K}  {\cal S}   & = {\bf 1} \cr
{\cal S}^t {\cal M}^2{\cal S}   & = {\cal M}^2_d\cr
\pmatrix{ W_\mu^3 \cr  B_\mu \cr  B'_\mu \cr}
 &= {\cal S}.\pmatrix {A_\mu \cr  Z_\mu \cr  Z'_\mu \cr}\cr}
}

To get a feeling for what we can expect as a function of the various
parameters, let us explore $\cal S$ in the limit of small $\varepsilon$. To
first order in $\varepsilon$, we can write:
\eqn\eIviii{\eqalign{ {\cal K}    & = {\bf 1} +
\varepsilon \ \Delta \cr {\cal S}    &    = {\cal O}_W^t {\cal O}_3^t({\bf
1} +
\varepsilon\ {\cal G}) \cr}
}
where
\eqn\eIix{{\cal O}_W = \pmatrix{s_W & c_W & 0 \cr
                      -c_W & s_W & 0 \cr
                                     0  &  0   & 1 \cr}
}
and
\eqn\eIx{{\cal O}_3 = \pmatrix{ 1 &     0 & 0    \cr
                       0 &   c_3 & -s_3  \cr
                       0 &  s_3 & c_3  \cr}
}
is the standard rotation matrix that describes the $Z$--$Z'$ mixing for
$\varepsilon = 0$. Then, from \eIvii{} we get two equations for ${\cal G}$:
\eqn\eIxi{\eqalign{
   {\cal G}  \ +  \ {\cal G}^t \ + \ \Delta'  & = 0  \cr
   {\cal M}^2_{d}\vert_{\varepsilon = 0}\ {\cal G} \  + \ \
   {\cal G}^t \ {\cal M}^2_{d}\vert_{\varepsilon = 0}
    & = {\cal M}^2_{d} - {\cal M}^2_{d}\vert_{\varepsilon = 0}, \cr}
}
with
\eqn\eIxii{ \Delta' = {\cal O}_3 {\cal O}_W \Delta {\cal O}_W^t {\cal
O}_3^t  =
{\pmatrix {
      0            &  - s_3 s_W              &   c_3 s_W \cr
  - s_3 s_W        &    2c_3 s_3c_W          & -(c_3^2 - s_3^2) c_W  \cr
    c_3 s_W        &  -  (c_3^2 - s_3^2) c_W & - 2c_3 s_3c_W
\cr}}.
}
The solution is:
\eqn\eIxiii{{\cal G} = {\pmatrix{
              0  &    s_3 s_W                 &   - c_3 s_W \cr
              0  & - c_3 s_3 c_W              & l(c_3^2 - s_3^2) c_W \cr
              0  &-(l-1)( c_3^2 - s_3^2) c_W  & c_3 s_3 c_W \cr}}
}
with $ l =  M_{Z'}^2 / (M_{Z'}^2 - M_Z^2) $.

The diagonalization of ${\cal M}^2$ yields the following masses for $Z$ and
$Z'$:
\eqn\eIIi{M_Z^2 = M_Z^2 \vert_{\varepsilon = 0} (1 - 2 c_3 s_3 c_W
\varepsilon)
}
\eqn\devmz{ M_{Z'}^2 = M_{Z'}^2 \vert_{\varepsilon = 0} (1 + 2 c_3 s_3 c_W
\varepsilon)
}
Notice that if $\theta_3 = 0$, then $ \Delta$  has zeros in the
diagonal  and from \eIxiii, ${\cal G}$  has the same property.
Then, from \eIxi\
\eqn\devmi{\delta(M_i^2) = \Sigma_j { \cal G }_{ij} M_j^2 \delta_{ji} = 0
}
and the masses are not corrected.\foot {This is a general result: when the
canonical kinetic term is changed without modifying the diagonal terms in the
base which diagonalizes the  gauge boson mass matrix, these masses  remain
unchanged at first order and only a redefinition of the fields is induced.}

The lesson is that the impact of  $\varepsilon$ on the eigenvalues is only of
order $\theta_3 \, \varepsilon$ in the limit of small $\varepsilon$. In
particular, the influence of ${\varepsilon}$ on the observable $M_Z$ may be
masked by a small value of $\theta_3$.

The value of $\theta_W$ is obtained by solving:
\eqn\valthet{\eqalign{
 (M^2_Z + M^2_{Z'}) \vert_{\varepsilon = 0}
& = {4 \pi \alpha \over s^2_W c^2_W}
  \left[ v^2 \left({1 \over 4}+{ 4 \over 25}\lambda^2\right)
    + \lambda^2 V^2\right] \cr
M^2_Z  M^2_{Z'} \vert_{\varepsilon = 0} & = { \pi \alpha \over s^2_W c^2_W}
                 \lambda^2 v^2 V^2 \cr }
}
for $v^2$. Taking into account that
\eqn\vev{v^2 = { 1 \over \sqrt{2} G_\mu}{1 \over 1 - \Delta r},
}
$\theta_W$ can be extracted as a function of the fundamental parameters
$\alpha, G_\mu, M_Z, M_{Z'}, \lambda, \varepsilon$ and of $\theta_3$ (although
this one is not independent in our model) and obeys:
\eqn\tanthet{\sin \theta_3 = { 4 \lambda D/5 \over
   \sqrt{(M^2_{Z'} - D)(M^2_{Z'} - M^2_Z)} \vert_{\varepsilon = 0} }
}
where
\eqn\valD{D =  { \pi \alpha \over s^2_W c^2_W}{ 1 \over \sqrt{2} G_\mu}
{1 \over 1 - \Delta r}.
}
Again, the variation of $\sin^2 \theta_W$ only probes the $\varepsilon
\theta_3$ combination:
\eqn\varsinW{\sin^2  \theta_W = \sin^2 \theta_W  \vert_{\varepsilon = 0}
\left( 1 - 2 \varepsilon  c_3s_3{ c_W^3 \over c^2_W - s^2_W}\right)
}

We have given above the corrections induced by a small $\varepsilon$. Another
interesting limit is to take $\lambda$ small relative to $\varepsilon$ and to
solve for the $\cal S$ matrix exactly in $\varepsilon$. The effects on the
low-energy observables would then be of order $\varepsilon \tilde \theta_3$
where $\tilde  \theta_3$ is a mixing angle of $O (\varepsilon M_Z^2 /
M_Z'^2)$.

If $\lambda$ and $\varepsilon$ are of the same order and the $Z'$ is not very
massive, the situation is more complicated and the matrix $S$ must be
evaluated numerically. If it is massive enough, a heavy $Z'$ decouples, as
both  $\theta_3$ and $\tilde \theta_3$ become negligible, and the matrix $\cal
S$  reduces to the very simple form
\eqn\simpS{{\cal S} = {\pmatrix {
              s_W  & -c_W & -\varepsilon \cr
              c_W  &  s_W & 0 \cr
              0  & 0  & 1 \cr}}
}
which will be useful in the following.

To study the process $e^+ e^- \rightarrow W^+ W^-$ we finally need the Feynman
rules for neutral currents and Three-Gauge-boson-Vertices (TGV).
Those are deduced from
\eqn\defi{{\cal L}_{NC} = ( J_{\mu}^3 \ \ J_{\mu}^Y \ \ J_{\mu}^{Y'} )
\cdot {\rm diag} ( g_2
\ \ g_1 \ \ g_{1'} ) \cdot {\pmatrix { W_3^{\mu} \cr B^{\mu} \cr B'^{\mu}
\cr } }
}
and
\eqn\defii{{\cal L}_{TGV} = - i g_V [({\hat W}_{\mu \nu}^{+} W^{\mu-} -
{\hat W}_{\mu \nu}^- W^{\mu +}) V^\nu
 + \kappa_V  W_{\mu}^+ W_\nu^-  V^{\mu \nu}]
}
where $ V = \gamma, Z, Z'$, and $\hat W_{\mu \nu}$ is defined from \eIvi.

{}From the definition of $\cal S$ and $O_{WB'}$ the couplings of the physical
gauge bosons are
\eqn\defiv{\eqalign{
(g^\gamma,\, g^Z, \,g^{Z'})
 & =(g_2,0,0) . {\cal S}\cr
\left( g^\gamma\kappa^\gamma, \,g^Z\kappa^Z,
\,g^{Z'}\kappa^{Z'}\right)
 & = (g_2, \,0,\,g_2 \varepsilon) .{\cal S}\cr
\left( e^\gamma_{L(R)},\,e^Z_{L(R)},\,e^{Z'}_{L(R)}\right)
 & =(g_2 T^3_{e_{L(R)}}, \, g_1 Y^B_{e_{L(R)}}, \,g'_1
      Y^{B'}_{e_{L(R)}}).{\cal S}\cr
e_{V(A)}
 &  = {1 \over 2}(e_R + (-) e_L), \cr }
}
following\foot{up to the correction of a trivial misprint} the notation
of \us.

The expression of the differential cross-section for the process $e^+ e^-
\rightarrow W^+ W^-$ can be found in \ref\WW{C.L. Bilchak and J.D. Stroughair,
\PRL {\bf D30}(1984) 1881.} and its generalization to the enlarged gauge group
$SU(2) \times U(1) \times U(1)'$  as a function of the previous parameters
given in \us{} is repeated here for the sake of completeness:
\eqn\defv{{ {d \sigma (e^+e^- \rightarrow W^+W^-)} \over {d \cos \theta} }
    = s^{1/2}(s/4 - M_W^2)^{1/2}  {\vert A \vert^2},
}
where $A$ takes into account the contributions from the four usual diagrams
corresponding to the $t$-channel $\nu$ exchange and the $s$-channel $\gamma$,
$Z$,  $Z'$ exchanges:
\eqn\defvi{\vert A \vert^2 = {1 \over 8 \pi}
    \sum_{\alpha,\beta} \left(
   a_{\alpha}^V\ Spin_{\alpha ,\beta}\ a_{\beta}^V +
   a_{\alpha}^A\ Spin_{\alpha ,\beta}\ a_{\beta}^A
                         \right),
}
with
\eqn\defvii{a^{V(A)}_{\alpha = \nu,\gamma, Z, Z'} =
\left( {  g^2_2  \over 4 t  },
       { e^\gamma_{V(A)} g^\gamma \over s},
       { e^Z_{V(A)} g^Z       \over s - M_Z^2},
       { e^{Z'}_{V(A)} g^{Z'} \over s - M_{Z'}^2} \right)
}
and
\eqn\defviii{\eqalign{
Spin_{\nu,\nu }
 &=\left[{ut \over {M_W^4} } -1 \right]
   \left[{t^2 \over 4 s^2} +{M_W^4 \over s^2}\right]
   + {t^2\over s M_W^2}\cr
Spin_{\nu,i}&=\left[{ut \over {M_W^4} } - 1 \right]
 \left[{ \kappa_i t\over 4 s} - {M_W^2 t\over 2s^2}
  - {M_W^4 \over s^2}\right]
  + (1+\kappa_i)\left[{ t \over 2M_W^2}
  - {t\over s} + {M_W^2 \over s}
 \right]\cr
Spin_{i,j}&=\left[{ut \over {M_W^4} } - 1 \right]
 \left[{ \kappa_i\kappa_j \over 4}
  - {M_W^2 \over s}{(1 + \kappa_i\kappa_j) \over 2}
  + 3{M_W^4 \over s^2}
 \right]\cr
&\ \ + (1+\kappa_i)(1+\kappa_j) \left[{s \over 4M_W^2}-1\right]
    ;\hskip2cm i=\gamma, Z, Z'\cr}
}
describing the angular dependence.

\newsec{Calculation of the Parameters ``\`a la Renard"}

The observability of the differential cross-section \defv\ was however only
discussed \us{} in terms of raw statistics, without considering, for example,
the explicit  angular resolution of the detectors. As future experiments are
often modelled  directly in terms of the final observable particles (leptons,
missing momenta  or jets) and related to fundamental Lagrangians in terms of
the Renard et al.  parameters,  {{\it e.g.} ($\delta g_Z$, $\delta \kappa_Z$,
$\delta  \kappa_{\gamma}$), we found it useful to provide the relation between
the  present approach and this common formulation. The apparent difficulty
here  stems from the fact that the Renard parametrization was written in full
generality only for the case of two neutral gauge bosons ($Z$ and $ \gamma$),
while we need here to take into account the $Z'$.

As we show below, it is fortunately possible ({\it for this particular
channel}) to reabsorb the $Z'$ exchange amplitude into a redefinition of the
$\gamma$- and $Z$-couplings to the $W$'s; the price to pay is that the
deviations from the standard model, $\delta g^Z$, $\delta \kappa^Z$ and
$\delta \kappa^\gamma$ now become $s$-dependent. Moreover, since $e^{Z'}_L$
and $e^{Z'}_R$ differ from their $Z$ counterparts, we will need to introduce a
non-zero $\delta g^\gamma$ as well. Notice that this effective parameter is
only a convenient effective way to mimic the presence of the $Z'$, and {\bf in
no way} implies a modification of the $W$'s electric charge.  {\it We think
however  that the necessity to introduce this extra parameter in the neutral
couplings  of $W$'s is interesting in itself and may be more general than the
present  framework. We therefore urge experimentalists to take this extra
freedom into  account in further studies.}

We now proceed as announced to integrate out the propagator of the $Z'$ into a
redefinition of the $Z$ and $\gamma$ couplings to the $W$'s. We  temporarily
affect these new couplings with a tilde; writing down the   different
contributions to the cross-section, we impose the conditions
\eqn\condi{\eqalign{
\sum_{j=\gamma,Z} {e_{L,R(SM)}^{\ j}\ \tilde g^j \over s - M^2_j}
  &=\sum_{i=\gamma,Z,Z'} {e_{L,R}^i g^i \over s -M^2_i}\cr
\sum_{j=\gamma,Z}{e_{L,R(SM)}^{\ j}\ \tilde g^j\tilde \kappa^j\over s - M^2_j}
  &= \sum_{i=\gamma,Z,Z'} {e_{L,R}^i g^i\kappa^i \over s -M^2_i}\cr}
}
where
\eqn\defix{\tilde \kappa^j = \kappa^j ; \, \tilde g^j = g^j
}
in the decoupling limit $M_{Z'} \rightarrow \infty$.

In addition to an explicit pole, the presence of the $Z'$ also affects the
mass and the couplings of the $Z$ by terms of order $\theta_3$, as we have
already seen. While we take these effects fully into account to determine  the
compatibility of the present model with LEP1 data, and for all the plots
reproduced below, we find it convenient to neglect  $\delta e_{L,R}^{Z}=
e_{L,R}^{Z}-e_{L,R(SM)}^{Z}$ (which is amply justified by the smallness of the
deviations, see the plots of $\delta e_{A,V}$ below) to provide more readable
analytic results. This allows us to rewrite the  equation \condi\ in the
following simple form
\eqn\seteq{ \eqalign{ \sum_{j=\gamma,Z} {e_{L,R}^{\
j}\ (\tilde g^j -
g^j) \over s -M^2_j} &=
 {e_{L,R}^{\ Z'}\ g^{Z'} \over s -M^2_{Z'}}\cr
\sum_{j=\gamma,Z} {e_{L,R}^{\ j}\ (\tilde g^j\tilde \kappa^j-
g^j\kappa^j)\over s -M^2_j} &=
 {e_{L,R}^{\ Z'}\ g^{Z'}\kappa^{Z'} \over s -M^2_{Z'}}\cr}
}

In the same approximation as above, and now introducing the convenient
notation $\delta g$ instead of $\tilde g$, we have:
\eqn\defdelta{\delta g^j\doteq\tilde g^j - g^j_{(SM)}
  \approx \tilde g^j - g^j
    + O(\varepsilon\theta_3, \varepsilon\tilde\theta_3)
}
(and similarly for $g^j \kappa^j$), thus parametrizing the effects of the $Z'$
as a departure from the SM's prediction. Solving what now are systems of two
equations with two unknowns yields
\eqn\resulti{\eqalign{\delta g^Z &=\,\, g^{Z'}\cdot {e_L^{Z'} - e_R^{Z'}
\over e_L^{Z} - e_R^{Z}}
\cdot{s - M_Z^2 \over s - M^2_{Z'}}\cr
\delta g^{\gamma} &= \,\,g^{Z'}\cdot{e_L^{Z}e_R^{Z'} - e_R^Z e_L^{Z'} \over
e^{\gamma}(e_L^{Z} - e_R^{Z})} \cdot
 {s  \over s - M^2_{Z'}}\cr}
}
and
\eqn\resultii{\eqalign{\delta (g^Z \kappa^Z) &= \,\,
g^{Z'}\kappa^{Z'}\cdot{e_L^{Z'} -
e_R^{Z'} \over e_L^{Z} - e_R^{Z}} \, \cdot {s - M_Z^2 \over s -
M^2_{Z'}}\cr
\delta (g^{\gamma} \kappa^{\gamma}) &=\,\,g^{Z'}
\kappa^{Z'}\cdot{e_L^{Z}e_R^{Z'} - e_R^Z
e_L^{Z'} \over e^{\gamma}(e_L^{Z} - e_R^{Z})} \cdot {s  \over s -
M^2_{Z'}}\cr}
}

{}From there on, solving for the $\cal S$ matrix in the appropriate limit  and
using  \defiv\ for the couplings yields the expected deviation from the SM for
$W$-pair production\foot{One should note that this parametrization is not
universal and is clearly not applicable as such, for example, to $W$-pair
production at hadron colliders or muons pairs production.}. For instance, the
simplest analytical expression for $\cal S$, {\it  i.e.} \simpS{}, in the
limit of decoupling $Z'$ ($M_Z' \gg M_Z$), gives:
\eqn\aproxdeli{\eqalign{\delta g^Z &= e \, \varepsilon\,\,({4 \lambda \over
5 s_W} -
\varepsilon\,{c_W \over s_W}) \cdot{s- M_Z^2 \over s -M_{Z'}^2} \cr
\delta g^{\gamma} &= e\, \varepsilon\,\,({\lambda \over 5 c_W^2 s_W}(1 + 4
s_W^2)  -
\varepsilon \,{s_W \over c_W}) \cdot{s  \over s -M_{Z'}^2}, \cr}
}
and in this limiting situation
\eqn\aproxdelii{\delta( g^Z \kappa^Z) =\delta(g^{\gamma}\kappa^{\gamma}) =0
}

Though being rather imprecise for the parameter region we are ultimately
interested in, \aproxdeli\ gives a good qualitative understanding  of the
effects of the $Z'$ and the extra anomalous coupling. It is interesting to
note that, in this limit, a non-vanishing $g^{Z'}$ persists to order
$\varepsilon$, while the $g^{Z'} \kappa^{Z'}$ vanishes. This is due to the
gauge invariance of the operator $O_{WB'}$. Had we used $\hat W_{\mu\nu}$
instead  of $ W_{\mu\nu}$ in \eIii{}, the fixed-angle, large-$s$ cross-section
would  have terms diverging like $s ({\kappa^{Z'} s \over M_{Z'}^2})^2$. These
cannot  be compensated by the $t$-channel which, in the decoupling limit, has
to  cancel the similar terms from the $\gamma$ and $Z$ exchanges: another
example  of gauge invariance enforcing a sound high-energy behaviour \deRuja.
This  observation persists to some extent in the non-decoupling situation, as
is  apparent from the numerical results below: the $\delta(g \kappa)$'s remain
much  smaller than the $\delta g$'s.

\newsec{LEP1 Constraints and Observability}

A clean way to check for the allowed range of our extra parameters (namely
$\lambda$ and $\varepsilon$, for any given value of $M_{Z'}$) is obviously to
proceed to a $\chi^2$ fit to the bulk of experimental data. Since the working
of such an analysis is unfortunately not  very easy to grasp intuitively, we
rather present a simpler exercise.  Namely, we will study here the impact of
$\lambda$ and $\varepsilon$ on the  most sensitive parameters of the Standard
Model, namely the vector and axial couplings ($e_A$, $e_V$) of the leptons to
the $Z$, as well as their reflection on the $W$ mass (we remind the reader
that we take the  standard option of fixing $M_Z$, $\alpha $ and the muon
decay constant  $G_{\mu}$). In the graphs presented below, we have (somewhat
abusively)  assumed that $\varepsilon = \lambda = 0$ corresponds to the
optimum Standard Model value, and compared the changes in $e_A$, $e_V$, $M_W$
to the  current error bar \cathy. The results are presented for $M_{Z'}=$ 210
or 300 GeV\foot{Notice that such low values of $M_Z'$ are not excluded,
provided $\lambda$ is small enough: for most $Z'$ models, $\lambda=O(1)$, and
masses smaller than 400 GeV are then excluded\ref\AHMQ{F. del Aguila, W.
Hollik, J.M. Moreno and M. Quir\'os, \NP{} {\bf B372} (1992) 3.}}  (see
\fig\eaev {Deviations from the Standard Model of low-energy observables
($M_W$, $e_A$ and $e_V$) for $M_{Z'} =$ 210 or 300~GeV/c$^2$. Different curves
correspond to different values of $\lambda$, ranging from 0 (plain curve) to
0.2 (smallest dashes), while the dependence on $\varepsilon$ can be followed
along a given curve. All curves here, and in what follows, are limited to the
portion where they satisfy the simultaneous constraints from LEP1 data and
$|\delta M_W|<0.52$ GeV/c$^2$, i.e. inside a 2$\sigma$ departure from current
values.}). As seen from those graphs, even for as  low a $Z'$ mass as 210 GeV,
$\varepsilon\in[-0.1, 0.1]$ is allowed for $\lambda \leq 0.1$. We have checked
that these conclusions are consistent with a more refined $\chi^2$ analysis.

Having established the LEP1-allowed range for our extra parameters, as a
function of $M_{Z'}$, we now proceed to plot the expected effects at
higher-energy colliders, taking as a standard example LEP2 at $\sqrt s =$
200~GeV. A typical plot of the differential cross-section can be found in \us.
Note that according to  the conventions of {\WW} we take $\theta$ to be the
emission angle of the  $W^+$ with respect ot the incoming $e^-$.  The
integrated cross-section suffers from cancellations between the  $\nu$--$Z'$
and  the ($\gamma, Z$)--$Z'$ interference terms, as expected from gauge
invariance. A decrease in the backward cross-section partly compensates an
increase in the  forward region, this compensation being exact around an angle
of 135$^\circ$, depending on the parameters.  It is therefore useful, if we
want to pursue the  analysis in terms of cross-sections, to define an
asymmetry, which crystallizes in one  single number this peculiar angular
dependence:
\eqn\asym{A_{135}\doteq
{\sigma_{\theta>135}-\sigma_{\theta<135}\over \sigma_{tot}.}
}
Since $\theta=135^\circ$ also happens to roughly equal the median of the
angular  distribution at 200 GeV, this asymmetry is only penalized by a factor
of $\sqrt2$ for its  statistical errors with respect to the total
cross-section, and yet it is most sensitive to the $Z'$ and $\varepsilon$
effects. On the basis of $10^4$  events, this corresponds to an optimal (\ie{}
statistical) 1$\sigma$-sensitivity  of 1.4\% for this asymmetry, and there is
thus room for 2$\sigma$ effects  within the range of LEP1-acceptable
parameters, even for $M_{Z'}$ as high as  300 GeV (see \us). For $M_{Z'}=$ 210
GeV, the statistical errors  are negligible  with respect to the deviations
from the SM, as can be seen in \fig\asymetry{The asymmetry $A_{135}$ as defined
in \asym for $\sqrt s=$200 GeV, $M_{Z'}=$ 210 GeV, as a function of
$\varepsilon$ and for $\lambda$ ranging from 0 (plain curve) to 0.2 (smallest
dashes). The SM would correspond to the maximum of the plain curve.}.

As we already mentioned, the differential cross-section is only very
indirectly accessible to experiment, and the reconstruction of the angular
distribution of $W$'s may prove very  difficult close to threshold. For that
reason, experimental groups often  prefer to rely on Monte-Carlo simulations
to relate directly effective  Lagrangian parameters to the actually observed
decay products (charged  leptons, jets or missing momenta). Particularly
useful for this purpose is a  general parametrization of the effective $W$
couplings developed by Renard and collaborators. This parametrization is in
fact exhaustive for the Standard  Model spectrum, with $\gamma$ and $Z$ the
sole neutral bosons exchanged. As we  saw above, we found it possible, and
hopefully useful for future experimental analysis, to rephrase the
extended model considered here (with an  explicit $Z'$) in terms of these
parameters. We found in fact that the effect  of the $Z'$, for $e^+ e^-
\rightarrow W^+ W^- \rightarrow final\, state$,  could be rephrased entirely
in terms of the $\delta g$, $\delta \kappa$  parameters with three small
caveats:
\item{--} the parametrization only applies to this channel ({\it not}
to $p$--$\bar{p}$  production, for instance);

\item{--} the coefficients $\delta g \, , \, \delta \kappa$ become in fact form
factors, depending upon $\sqrt s$ (which, if we neglect initial state
radiation, is in practice fixed);

\item{--} we need to introduce an extra parameter $\delta g_\gamma$; we insist
again on the fact that $\delta g_\gamma$ is {\bf not} to be interpreted as a
change in the $W$ electric charge, but only as a convenient way to simulate
part of the $Z'$ exchange.

The graphs are fairly self-explanatory, the various lines have been
deliberately limited to plot only for values of $\lambda$ and $\epsilon$
leading to less than 2 $\sigma$ departures in \fig\dgs{The relative deviations
from the SM triple-gauge couplings induced at $\sqrt s=200$~GeV by a $Z'$ of
210~GeV, as a function of its anomalous coupling ($\varepsilon$), and for a
gauge coupling ranging from $\lambda=0$ (plain curve) to $\lambda=0.2$ (small
dots). Figures for other values of $\sqrt s$ and $M_{Z'}$ can be obtained to a
very good accuracy by a simple rescaling according to the dependence manifest
in \resulti, \resultii.} and show in  particular that sizeable contributions
to $\delta g_Z$ and $\delta g_\gamma$  are allowed, while $\delta (g_Z
\kappa_Z)$ and $\delta (g_\gamma  \kappa_\gamma)$ are minute.  Of particular
interest is the $\delta g_Z$--$\delta g_\gamma$ plane (\fig\dggdgz{The
deviations $\delta g^Z/g_Z$ (horizontally) versus $\delta g^\gamma/g_\gamma$
(vertically). As previously, a curve corresponds to a continuous change in the
value of $\varepsilon$, while increasingly dotted lines correspond to
$\lambda$ increasing from 0 to 0.2.}), and we think this could be used quite
generally to  discuss departures from the Standard Model (for instance, the
present model  introduces a strong correlation between $\delta g_Z$ and
$\delta g_\gamma$,  while non-extended cases are confined to the horizontal
axis).

\newsec{Conclusions}

The purpose of the present paper was to investigate to what extent the severe
constraints, imposed by LEP1 and the assumption of $SU(2) \times U(1)$ gauge
invariance, on the possible observation of anomalous $W$ couplings at future
colliders could be alleviated. We have addressed one of the deeper assumptions
made in this type of approach, namely we have shown that a {\it larger}
symmetry group could in fact lead to {\it less } constraining conditions than
the minimal $SU(2) \times U(1)$. This apparent paradox stems from the fact
that the linear realization of the larger group requires a larger light-mass
spectrum than the minimal one usually assumed. We have only tried to produce
an  explicit example, involving the minimal extension of the gauge group to
$SU(2)  \times U(1) \times U(1)'$, and our results do not attempt at
generality (for  instance, we have not included all the obviously possible
extra effective  terms. This example however
allows us to  state our point, namely that departures of the standard
couplings larger than  those expected for the minimal symmetry are indeed
possible, provided some  extra bosons are light, though weakly coupled. Such
small values of the gauge  couplings make the presence of $Z'$ bosons even as
light as 210 GeV perfectly  compatible with current results from hadronic
colliders  \ref\cdf{CDF collaboration, F. Abe {\it et al} \PRL {\bf 65}, 2243
(1990)}.

In the process of this analysis, we have been brought to formulate our results
in terms of the Renard et al.  parameters  $\delta g_Z$,
$\delta\kappa_{Z,\gamma}$, and we have shown that these parameters could be
easily generalized to mimic in a specific channel the contribution of $Z'$, at
the cost of adding $\delta g_{\gamma}$ to their number. We have also suggested
the use of correlated plots in the space $\delta g_{\gamma}$, $\delta g_Z$ to
discuss the results or expectations of the various experiments. For some values
of the parameters, other channels might reveal the presence of $Z'$ earlier or
even more strikingly: for instance, larger  values of its direct coupling to
leptons (governed by $\lambda$) might favour the $\mu$--$\bar \mu $ channel.
This depends in each case on the set of parameters and of the experimental
set-up; in any case such observability  detracts in no way from the present
conclusions.

\noindent {\bf Acknowledgements}\par\nobreak\medskip\nobreak
The material in this paper has been the subject of numerous discussions with
colleagues, either theorists or experimentalists, at CERN and during the
Moriond Meeting. In particular, Alvaro De R\'ujula, Bel\'en Gavela and
Olivier P\`ene, Misha Vysotsky, Georges Gounaris, E. Mass\'o, M.
Mart\'{\i}nez, S. Peris, Alain Blondel, Stavros Katsanevas, Daniel Treille and
the  DELPHI LEP200 group.

\listrefs
\listfigs
\vfill\supereject
\vbox {\offinterlineskip
\hrule
\halign
{&\vrule#&\strut\quad\hfil#&\vrule#&\strut\quad
\hfil#\quad&\vrule#&\strut\quad\hfil#\quad
&\vrule#&\strut\quad\hfil#\quad&\vrule#&\strut\quad
\hfil#\quad&\vrule#&\strut\quad\hfil#\quad&\vrule#
&\strut\quad\hfil#\quad&\vrule#&\strut\quad\hfil#\quad\cr
height2pt&\omit&&\omit&&\omit&&\omit&&\omit&&\omit&&\omit&&\omit&&\omit&\cr
& Hypercharge \hfil && ${\pmatrix{u\cr  d\cr}}_L$ && $u_L^c$ &&
$d_L^c$ && ${\pmatrix{\nu\cr e\cr}}_L$ && $e_L^c$ && $\nu_L^c$ &&
$\pmatrix{\phi_1\cr \phi_2\cr}$ && $\chi$\ \ &\cr
height2pt&\omit&&\omit&&\omit&&\omit&&\omit&&\omit&&\omit&&\omit&&\omit&\cr
\noalign{\hrule}
height2pt&\omit&&\omit&&\omit&&\omit&&\omit&&\omit&&\omit&&\omit&&\omit&\cr
 & \ \ && \ \ && \ \ && \ \ && \ \ && \ \ && \ \ && \ \ && \ \ & \cr &
$Y$ \ \ \ \ \ \ && ${1\over 6}$ \ \ \ \ && $ -{2\over 3}$ && ${1\over
3}$ && $-{1\over 2}$ \ \  && $\scriptstyle1$ && $\scriptstyle 0$ &&
${1\over 2}$ \ \ && $\scriptstyle 0$\ \ &\cr & \ \ && \ \ && \ \ && \
\ && \ \ && \ \ && \ \ && \ \ && \ \ & \cr
& $Y'$ \ \ \ \ \ \ &&
${1\over5}$ \ \ \ \ && $ {1\over 5}$ && $ -{3\over 5}$ && $ -{3\over
5}$ \ \  &&  ${1\over5}$ && $\scriptstyle1$ && $-{2\over 5}$ \ \ &&
$\scriptstyle1$\ \ &\cr & \ \ && \ \ && \ \ && \ \ && \ \ && \ \ && \
\ && \ \ && \ \ & \cr height2pt
&\omit&&\omit&&\omit&&\omit&&\omit&&\omit&&\omit&&\omit&&\omit&\cr}
\hrule}
\vskip 0.5in

Table 1: Matter content and charges assignment of the extended
model $SU(2)\times U(1)\times U(1)^\prime$.}

\vfill\supereject
\null\vskip1.5cm
\line{
 \psannotate{\psboxto(0.48\hsize;0cm){dmw210.eps}}
  {%\fillinggrid%
   \at(10\pscm;0.4\pscm){$\varepsilon$}%
   \at(5.4\pscm; 6.5\pscm){\hbox to 0pt{\hss
      $\delta M_W(\varepsilon)$; $M_{Z'}=210$ GeV\hss}}}
 \hfil
 \psannotate{\psboxto(0.48\hsize;0cm){dmw300.eps}}
  {%\fillinggrid%
   \at(10\pscm;0.4\pscm){$\varepsilon$}%
   \at(5.4\pscm; 6.4\pscm){\hbox to 0pt{\hss
      $\delta M_W(\varepsilon)$; $M_{Z'}=300$ GeV\hss}}}
}
\goodbreak
\null\vskip1cm
\line{
 \psannotate{\psboxto(0.48\hsize;0cm){deaz210.eps}}
  {%\fillinggrid%
   \at(10\pscm;0.4\pscm){$\varepsilon$}%
   \at(5.4\pscm; 6.2\pscm){\hbox to 0pt{\hss
      $\delta e_A^Z(\varepsilon)$; $M_{Z'}=210$ GeV\hss}}}
 \hfil
 \psannotate{\psboxto(0.48\hsize;0cm){deaz300.eps}}
  {%\fillinggrid%
   \at(10\pscm;0.4\pscm){$\varepsilon$}%
   \at(5.4\pscm; 6.2\pscm){\hbox to 0pt{\hss
      $\delta e_A^Z(\varepsilon)$; $M_{Z'}=300$ GeV\hss}}}
}
\goodbreak
\null
\line{
 \psannotate{\psboxto(0.48\hsize;0cm){devz210.eps}}
  {%\fillinggrid%
   \at(10\pscm;0.4\pscm){$\varepsilon$}%
   \at(5.4\pscm; 6.2\pscm){\hbox to 0pt{\hss
      $\delta e_V^Z(\varepsilon)$; $M_{Z'}=210$ GeV\hss}}}
 \hfil
 \psannotate{\psboxto(0.48\hsize;0cm){devz300.eps}}
  {%\fillinggrid%
   \at(10\pscm;0.4\pscm){$\varepsilon$}%
   \at(5.4\pscm; 6.2\pscm){\hbox to 0pt{\hss
      $\delta e_V^Z(\varepsilon)$; $M_{Z'}=300$ GeV\hss}}}
}
\vskip0.3cm
\item{ } Fig. 1: Deviations from the Standard Model of low-energy observables
($M_W$, $e_A$ and $e_V$) for $M_{Z'} =$ 210 or 300~GeV/c$^2$. Different curves
correspond to different values of $\lambda$, ranging from 0 (plain curve) to
0.2 (smallest dashes), while the dependence on $\varepsilon$ can be followed
along a given curve. All curves here, and in what follows, are limited to the
portion where they satisfy the simultaneous constraints from LEP1 data and
$|\delta M_W|<0.52$ GeV/c$^2$, i.e. inside a 2$\sigma$ departure from current
values.

\vfill\supereject
\null\vskip1.5cm
\psannotate{\psboxto(\hsize;0cm){asy210.eps}}
  {%\fillinggrid%
   \at(10\pscm;0.4\pscm){$\varepsilon$}%
   \at(5.4\pscm; 6.2\pscm){\hbox to 0pt{\hss
      $A_{135}(\varepsilon)$;  $M_{Z'}=210$ GeV\hss}}}
\vskip 1cm
\item{ } Fig. 2: The asymmetry $A_{135}$ as defined
in \asym for $\sqrt s=$200 GeV, $M_{Z'}=$ 210 GeV, as a function of
$\varepsilon$ and for $\lambda$ ranging from 0 (plain curve) to 0.2 (smallest
dashes). The SM would correspond to the maximum of the plain curve.

\vfill\supereject
\null\vskip1.5cm
\line{
 \psannotate{\psboxto(0.48\hsize;0cm){dgg210.eps}}
  {%\fillinggrid%
   \at(10\pscm;0.4\pscm){$\varepsilon$}%
   \at(5.4\pscm; 6.2\pscm){\hbox to 0pt{\hss
      $\delta g^\gamma(\varepsilon)/g^\gamma$; $M_{Z'}=210$ GeV\hss}}}
 \hfil
 \psannotate{\psboxto(0.48\hsize;0cm){dgz210.eps}}
  {%\fillinggrid%
   \at(10\pscm;0.4\pscm){$\varepsilon$}%
   \at(5.4\pscm; 6.2\pscm){\hbox to 0pt{\hss
      $\delta g^Z(\varepsilon)/g^Z$; $M_{Z'}=210$ GeV\hss}}}
}
\null\vskip1cm
\line{
 \psannotate{\psboxto(0.48\hsize;0cm){dgkg210.eps}}
  {%\fillinggrid%
   \at(10\pscm;0.4\pscm){$\varepsilon$}%
   \at(5.7\pscm; 6.2\pscm){\hbox to 0pt{\hss
      $\delta (g^\gamma\kappa^\gamma)(\varepsilon)/g^\gamma\kappa^\gamma$;
       $M_{Z'}=210$ GeV\hss}}}
 \hfil
 \psannotate{\psboxto(0.48\hsize;0cm){dgkz210.eps}}
  {%\fillinggrid%
   \at(10\pscm;0.4\pscm){$\varepsilon$}%
   \at(5.7\pscm; 6.2\pscm){\hbox to 0pt{\hss
      $\delta (g^Z\kappa^Z)(\varepsilon)/g^Z\kappa^Z$; $M_{Z'}=210$ GeV\hss}}}
}
\vskip0.5cm
\item{ } Fig. 3: The relative deviations
from the SM triple-gauge couplings induced at $\sqrt s=200$~GeV by a $Z'$ of
210~GeV, as a function of its anomalous coupling ($\varepsilon$), and for a
gauge coupling ranging from $\lambda=0$ (plain curve) to $\lambda=0.2$ (small
dots). Figures for other values of $\sqrt s$ and $M_{Z'}$ can be obtained to a
very good accuracy by a simple rescaling according to the dependence manifest
in \resulti, \resultii.

\vfill\supereject
\null\vskip1cm
\psannotate{\psboxto(\hsize;0cm){dgzdgg210.eps}}
  {%\fillinggrid%
   \at(10\pscm;0.4\pscm){$\delta g^Z$}%
   \at(1.1\pscm; 6.2\pscm){$\delta g^\gamma$}}

\vskip0.5cm
\item{ } Fig. 4: The deviations $\delta g^Z/g_Z$ (horizontally) versus $\delta
g^\gamma/g_\gamma$ (vertically). As previously, a curve corresponds to a
continuous change in the value of $\varepsilon$, while increasingly dotted
lines correspond to $\lambda$ increasing from 0 to 0.2.
\autojoin
\bye